\documentclass[journal]{IEEEtran}
\ifCLASSINFOpdf
\else
\fi
\usepackage{graphicx}  
\usepackage{bm}        
\usepackage{amssymb}   
\usepackage{amsmath}
\usepackage{float}
\usepackage{color}
\usepackage{physics}

\begin{document}
%
\title{Robust optical frequency transfer in a noisy urban fiber network}
%
%
%
\author{Xiang~Zhang, Liang~Hu,~\IEEEmembership{Member,~IEEE,}  Xue~Deng, Qi~Zang, Jie~Liu, Dongjie Wang, Tao~Liu, Ruifang~Dong, and~Shougang~Zhang
\thanks{This work was supported in part by the National Key Research and Development Program of China (Grant No. 2016YFF0200200), in part by the National Natural Science Foundation of China (NSFC) (91636101, 91836301 and 11803041), in part by the West Light Foundation of the Chinese Academy of Sciences (Grant No. XAB2016B47), and in part by the Strategic Priority Research Program of the Chinese Academy of Sciences (Grant No. XDB21000000).\emph{Corresponding author: Tao Liu and Ruifang Dong}}

\thanks{X. Zhang, X. Deng, Q. Zang, D. Wang, Q. Zhou, J. Liu, D. Jiao, J. Gao, D. Wang, T. Liu, R. Dong, S. Zhang are with National Time Service Center, Chinese Academy of Sciences, Xi'an 710600, China; University of Chinese Academy of Sciences, Beijing 100049, China; Key Laboratory of Time and Frequency Standards, Chinese Academy of Sciences, Xi'an 710600, China (e-mail: zhangxiang@ntsc.ac.cn; dengxue@ntsc.ac.cn; taoliu@ntsc.ac.cn; dongruifang@ntsc.ac.cn).}
\thanks{L. Hu is with the State Key Laboratory of Advanced Optical Communication Systems and Networks, Shanghai Institute for Advanced Communication and Data Science, Shanghai Key Laboratory of Navigation and Location-Based Services Department of Electronic Engineering, Shanghai Jiao Tong University, Shanghai 200240, China (e-mail: liang.hu@sjtu.edu.cn).}}
%
%

\markboth{JOURNAL OF LIGHTWAVE TECHNOLOGY,~Vol.~xxx, No.~xxx, xxx~2021}%
{Shell \MakeLowercase{\textit{et al.}}: Bare Demo of IEEEtran.cls for IEEE Journals}
%



\maketitle

\begin{abstract}
Optical fibers have been recognized as one of the most promising host material for high phase coherence optical frequency transfer over thousands of kilometers. In the pioneering work, the active phase noise cancellation (ANC) technique has been widely used for suppressing the fiber phase noise introduced by the environmental perturbations, in which an ideal phase detector with high resolution and unlimited detection range is needed to extract the fiber phase noise, in particular for noisy fiber links. We demonstrate the passive phase noise cancellation (PNC)  technique without the need of phase detector could be preferable for noisy fiber links. To avoid the effect of the radio frequency (RF) from the time base at the local site in the conventional active or passive phase noise cancellation techniques, here we introduce a fiber-pigtailed acousto-optic modulator (AOM) with two diffraction order outputs (0 and +1 order) with properly allocating the AOM-driving frequencies allowing to cancel the time base effect. Using this technique, we demonstrate transfer of coherent light through a 260 km noisy urban fiber link. The results show the effect of the RF reference can be successfully removed. After being passively compensated, {we demonstrate a fractional frequency instability of $4.9\times10^{-14}$ at the integration time of 1 s and scales down to $10^{-20}$ level at 10,000 s in terms of modified Allan deviation over the 260 km noisy urban fiber link}. The frequency uncertainty of the retrieved light after transferring through this noise-compensated fiber link relative to that of the input light achieves $(0.41\pm4.7)\times10^{-18}$. The proposed technique opens a way to a broad distribution of an ultrastable frequency reference with high coherence without any effects coming from the RF reference and enables a wide range of applications beyond metrology over fiber networks.
\end{abstract}


%
\IEEEpeerreviewmaketitle

\section{Introduction}



\begin{figure*}[htbp]
\centering
\includegraphics[width=1\linewidth]{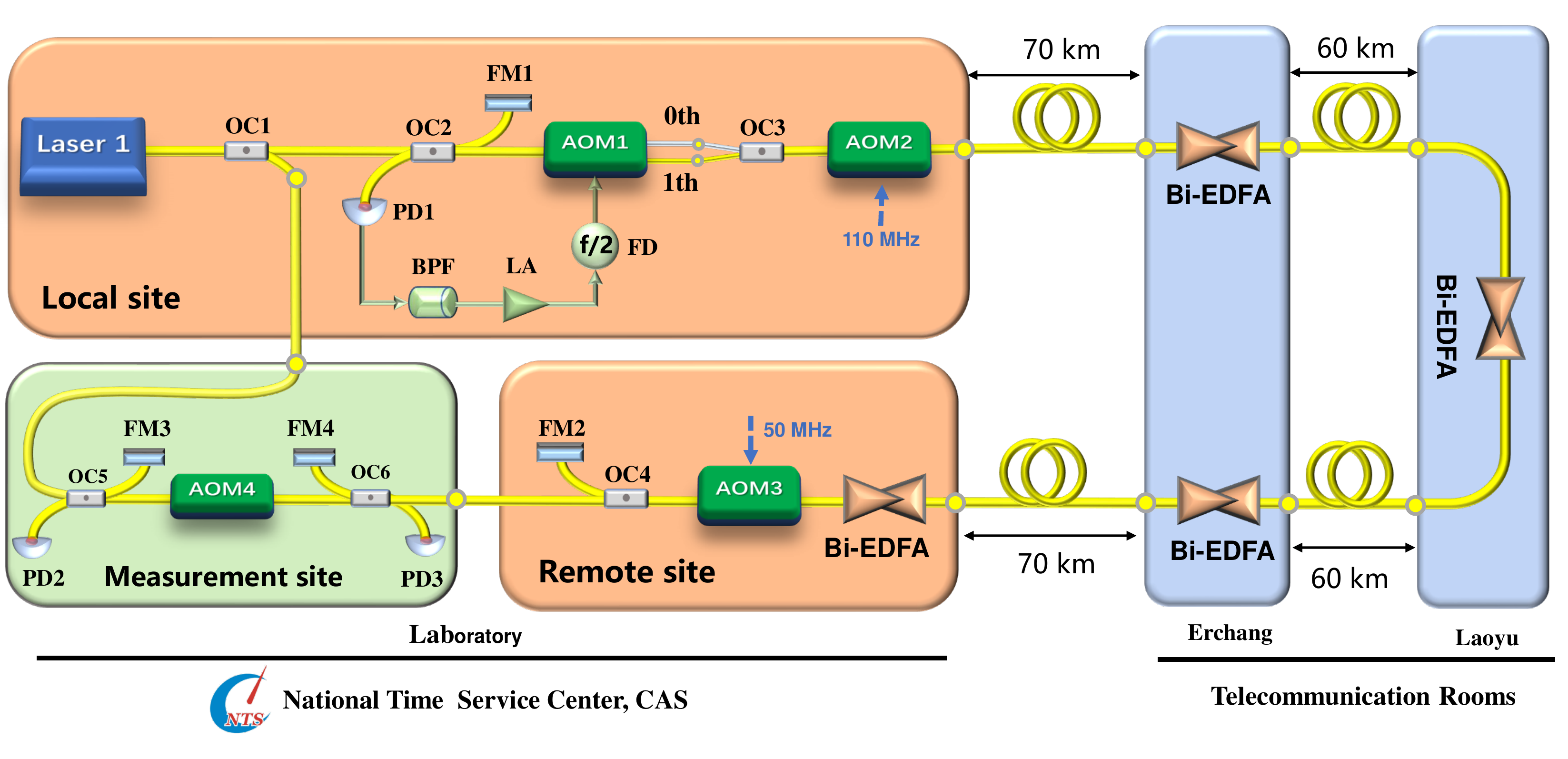}
\caption{Schematic  of optical frequency transfer via a 260km urban fiber link by adopting the  passive phase noise cancellation technique without the effect of the time base. AOM: acousto-optic modulator, FM: Faraday mirror, Bi-EDFA: bi-directional erbium-doped fiber amplifier, OC: optical coupler, FD: frequency divider, PD: photo-detector.}
\label{fig1}
\end{figure*}

\IEEEPARstart{O}{ptical}  frequency timekeeping with trapped atoms or ions \cite{PhysRevLett.120.103201, schioppo2017ultrastable, McGrew:2018aa} is progressively becoming ideal tools for precision measurements and fundamental physics tests, such as general relativity\cite{chou2010,delva2017}, temporal variation of the fundamental constant \cite{parker2018measurement}, searching for dark matter\cite{Roberts_2020}, chronometric geodesy \cite{grotti2018geodesy}, and gravitational waves \cite{kolkowitz2016gravitational, graham2013new, hu2017atom}. Unfortunately, these clocks are typically cumbersome and expensive and only available at national metrology institutes and several universities \cite{ PhysRevLett.120.103201, schioppo2017ultrastable, takamoto2005optical}, causing a strong motivation to develop a robust and cost-effective system for comparing and distributing these sources of ultraprecise frequency signals. Among different frequency transfer methods,  optical fiber links have been widely demonstrated as {one of the most powerful tools} for the dissemination and comparison of current state-of-the-art optical clocks with the stability and uncertainties at a few $10^{-21}$ level \cite{ma1994delivering, droste2013optical, calonico2014high}. However, the temporal optical length variations of the fiber link, coming from mechanical and thermal perturbations along the fiber link, will imprint optical phase noise onto the transmitted optical signal and ultimately deteriorate the optical frequency transfer stability and uncertainty. One of the most pioneering technique, namely active phase noise cancellation (ANC), has been widely adopted for mitigating this fiber-induced Doppler noise, where the phase noise of the fiber link is detected by comparing the local signal with the round-trip one and be used to implement a feedback control onto the phase compensator \cite{ma1994delivering}. Nevertheless, there exist obvious and inherent disadvantages for this ANC technique that it requires the complicated phase noise detection unit and the sophisticated servo controller. The overall performance of the phase noise rejection capability is strongly  dependent on the proper setting of the loop parameters. At the same time, an ideal phase detector with high resolution and unlimited detection range is needed to extract the fiber phase noise. One common way to extend the phase detection range is to adopt an analog phase detector with an assistance of a high frequency dividing ratio \cite{Chiodo:15} or a digital phase detector with the proper phase unwrap algorithm \cite{redpitaya2018}. Consequently, a complicated phase noise detection circuit has to be used in a noisy fiber link with a large phase noise per-unit-length such as $4$ rad$^2$/Hz/km (USA) \cite{williams2008high}, $0.5$ rad$^2$/Hz/km (Germany) \cite{raupach2014optical}, $3$ rad$^2$/Hz/km (France) \cite{Chiodo:15, lopez2010cascaded}, $20$ rad$^2$/Hz/km (Italy) \cite{calonico2014high}, $20$ rad$^2$/Hz/km (Japan) \cite{akatsuka2020optical} at the 1 Hz offset frequency. The phase noise is even larger in China.  For example, the phase noise per-unit-length at the 1Hz is about $38.5$ rad$^2$/Hz/km for a 260 km urban fiber link connecting LinTong District and HuYi District and the integrated phase from 0.1Hz to 100 kHz are approaching to 240 rad (see the following text). For an analog phase detector, a frequency divider with a ratio of more than 240 is needed to extract the phase noise introduced by the fiber link. To secure the enough phase detection range caused by the fluctuations of the fiber link phase noise, the frequency divider with the larger frequency divider is needed. This could still not be enough for the phase noise of the fiber link fluctuated by more than one order of magnitude, resulting in the appearance of the cycle slips \cite{Chiodo:15}. Moreover, before adopting the frequency divider with the high dividing ratio, the beatnote signal at the local site typically needs to be up-converted as implemented in \cite{Chiodo:15}. This process will introduce additional phase noise coming from the frequency mixer and amplifier stages.

To surmount the aforementioned difficulties, Hu. \textit{et al} first proposed a passive phase noise cancellation (PNC) technique for optical frequency transfer by detecting the phase noise with a pilot optical signal and pre-compensating the same amount phase noise on the other forward optical signal, resulting in the phase noise compensated light automatically received at the remote site \cite{hu2020passive1, hu2021all}. They  demonstrated a proof-of-principle optical frequency transmission experiment via a 145 km spooled fiber link with a stability of $2\times10^{-19}$  at 10,000 s in terms of overlapping Allan deviation (OADEV) \cite{hu2020passive1}. Although the PNC technique has been demonstrated with the spooled fiber link \cite{hu2020passive1}, the capability for the phase noise rejection in the field-deployed fiber links needs to be experimentally examined as widely done by the ANC technique \cite{droste2013optical, calonico2014high,williams2008high, Lopez:12,Chiodo:15,deng2016}.

Neither the active nor the passive phase compensation schemes can avoid the influence of the time base at the local site, because for a practical application, the radio frequency (RF) signals synthesized from their time bases are independent of the optical frequency reference at the local site. By taking the effect of the time base into consideration, the limitation of the optical frequency transfer stability $\sigma_{y}$ can be expressed as.
\begin{equation}
\sigma_y\approx\frac{\omega_l}{2\omega_s}\sigma_{\text{RF}}
\end{equation}
where $\omega_l$ represents the driving frequency for the local acousto-optic modulator (AOM), and $\sigma_{\text{RF}}$  is the stability of the time base.  As our previous work, we have $\omega_l=2\pi\times 110$ MHz and $\omega_s =2\pi\times193$ THz. {For a typical radio frequency (RF) reference such as a customized OCXO with the stabilities of $8.7\times 10^{-11}$ at 1s and $7.2\times 10^{-11}$ at 10,000 s, it will constrain $\sigma_{y}$ of $5.0\times 10^{-17}$ at 1 s and $4.1\times 10^{-17}$ at 10,000 s}, illustrating that the long-term stability will become the limitation.  One way to solve this issue is to discipline the RF reference to the GPS signal \cite{cheng2005}, but this will increase the complexity of the system and the system performance is strongly dependent on the GPS performance. Therefore, one intriguing question is how  to distribute the optical signal without the effect of the time base. In a pioneering work, Wu \textit{et al} demonstrated the effect of the time base can be effectively removed by cascading one more AOM just after the remote AOM for optical frequency transfer over a branching fiber network\cite{Wu16}. However, adding one more AOM, in particular a fiber-pigtailed one, will introduce the significant outside path, resulting in the increase of the outside phase noise as discussed in \cite{stefani2015tackling, hu2020multinode, hu2021branching}. Additionally, the scheme still couldn't get rid of the effect of the time base at the local site.

 In this paper, a fiber-pigtailed acousto-optic modulator (AOM) with two diffraction order outputs ($0$ and $+1$ order) is employed as the phase noise compensated device, the retrieved optical frequency signal is independent of the time base at the local site. Furthermore, we demonstrate the PNC technique has unlimited phase noise detection range and resolution and could be preferred for noisy fiber links. We for the first time demonstrate the proposed technique by coherently transferring an ultra-stable optical frequency signal via a 260 km noisy urban fiber link.

The article is organized as follows. We illustrate the concept of coherent optical phase dissemination with the PNC technique without the effect of the time-base in Sec. \ref{sec2}. We present the experimental set-up and experimental results in Sec. \ref{sec4}. Finally, we give a conclusion in Sec. \ref{sec6}.

\section{Concept of cancelling the time base effect in an optical frequency transfer network}
\label{sec2}

Figure \ref{fig1} illustrates a schematic diagram of optical frequency transfer via a fiber with the PNC technique. Part of this light source is extracted as the local reference, while the remainding part is injected into an acousto-optic modulator (AOM1) with two diffracted outputs (0 and -1 order). The laser outputs from the two fiber-pigtails are further combined and coupled to the AOM2 working at the angular frequency of $\omega_l$ (upshifted mode, $+1$ order) and then propagated to the remote site via the fiber link. At the remote site, another AOM3 (upshifted mode, $+1$ order) working at the angular frequency of $\omega_r$ is used to discriminate the desired optical signal from the stray reflections by the fiber connectors and splices. The light is retro-reflected back to the local site and the phase noise of the fiber link is recovered by the photodetector 1 (PD1) at the local site. Subsequently, the frequency of the detected signal is divided by a factor of 2 and then drives the AOM1 with the -1 diffraction order. At the remote site, the phase noise is automatically cancelled at the remote site without the effect of the time base at the local site. The detailed principle of optical frequency transfer is as follows. The initial optical frequency signal can be denoted as,
\begin{equation}
E_1\propto\cos[\omega_st+\phi_s],
\end{equation}
where $\omega_s$ and $\phi_s$ respectively represent the angular frequency and initial phase value of the optical signal. Assuming the optical fiber phase noise induced by environmental perturbations in either direction is equal, the electric field of the light which is reflected by the Faraday mirror 2 (FM2) and transferred back to the local site  can be written as,
\begin{equation}
E_2\propto\cos[(\omega_s+2\omega_l+2\omega_r)t+\phi_s+2(\phi_l+\phi_f+\phi_r)],
\end{equation}
where $\phi_l$ and $\phi_r$, respectively, represent the initial phases of the driving frequencies of the AOM2 and AOM3, and the $\phi_f$ is the optical fiber induced phase noise. At the local site, the heterodyne beat-note between  $E_1$ and  $E_2$ is detected by the PD1. We select the electrical signal with the angular frequency of $2(\omega_l+\omega_r)$, which contains the fiber phase noise information, by using a narrow bandpass RF filter, resulting in,
 \begin{equation}
E_3\propto\cos[2(\omega_l+\omega_r)t+2(\phi_l+\phi_f+\phi_r)].
\label{eq4}
\end{equation}

Afterwards, the angular frequency of $E_3$ is divided by a factor of 2, obtaining,
 \begin{equation}
E_4\propto\cos[(\omega_l+\omega_r)t+(\phi_l+\phi_f+\phi_r)].
\end{equation}

To passively compensate the phase noise along the fiber link, we feed the amplified electrical signal $E_4$ to the AOM1. The phase noise compensated light output at remote site can be expressed as,
 \begin{equation}
E_5\propto\cos[\omega_st+\phi_s].
\end{equation}

As can be seen, the frequency and phase of the retrieved optical signal are independent of all the RF signals. Consequently, the RF signal source with the high frequency stability in our scheme is no longer needed, which will simplify the system and make it cost-effective. It is important to emphasize that the above description does not take the fiber propagation delay into the account. As demonstrated by Williams \textit{et al.},  the phase noise rejection capability is limited by the propagation delay \cite{williams2008high}. The residual phase noise power spectral density (PSD), $S_{\text{o}}(f)$, at the remote site in terms of the single-pass free-running phase noise PSD, $S_{\text{fiber}}(f)$, and the single-pass fiber link $L$ propagation delay, $\tau_0$, can be expressed  as \cite{williams2008high,hu2020passive1,hu2020multinode},
\begin{equation}
S_{\text{o}}(f)\simeq\frac{7}{3}{(2\pi f\tau_0)^2}S_{\text{fiber}}(f).
\label{eq7}
\end{equation}

\section{Experimental setup and results}
\label{sec4}

\subsection{Experimental setup}

We have demonstrated the proposed technique by using the configuration as shown in Fig. \ref{fig1}. The interferometer is built with fiber optics. A laser source, which is stabilized to  a high-finesses Fabry-Perot cavity at a frequency near 193 THz is adopted as an optical reference.  The optical fiber is composed of a pair of 130 km fiber link connecting the NTSC (National Time Service Center) at Lintong and two communication rooms of CTCC (China Telecom), with a total optical attenuation of 75 dB. To compensate the optical loss, four bi-directional erbium-doped fiber amplifiers (Bi-EDFAs) are independently installed along this fiber link.  In this experiment, we set $\omega_l = 2\pi\times 110$ MHz and $\omega_r = 2\pi\times 50$ MHz.  In order to conveniently evaluate the performance of this coherent frequency transmission system, a two-way optical frequency comparison setup is adopted as described in \cite{hu2021branching}. We use non-averaging $\Pi$-type frequency counters to record the beating frequency. Additionally, to measure the phase noise of optical frequency transfer, a commercial phase noise analyzer (Rohde $\&$ Schwarz, FSWP) is used.

\subsection{Residual phase noise PSDs at the remote site}

\begin{figure}[htbp]
\centering
\includegraphics[width=0.95\linewidth]{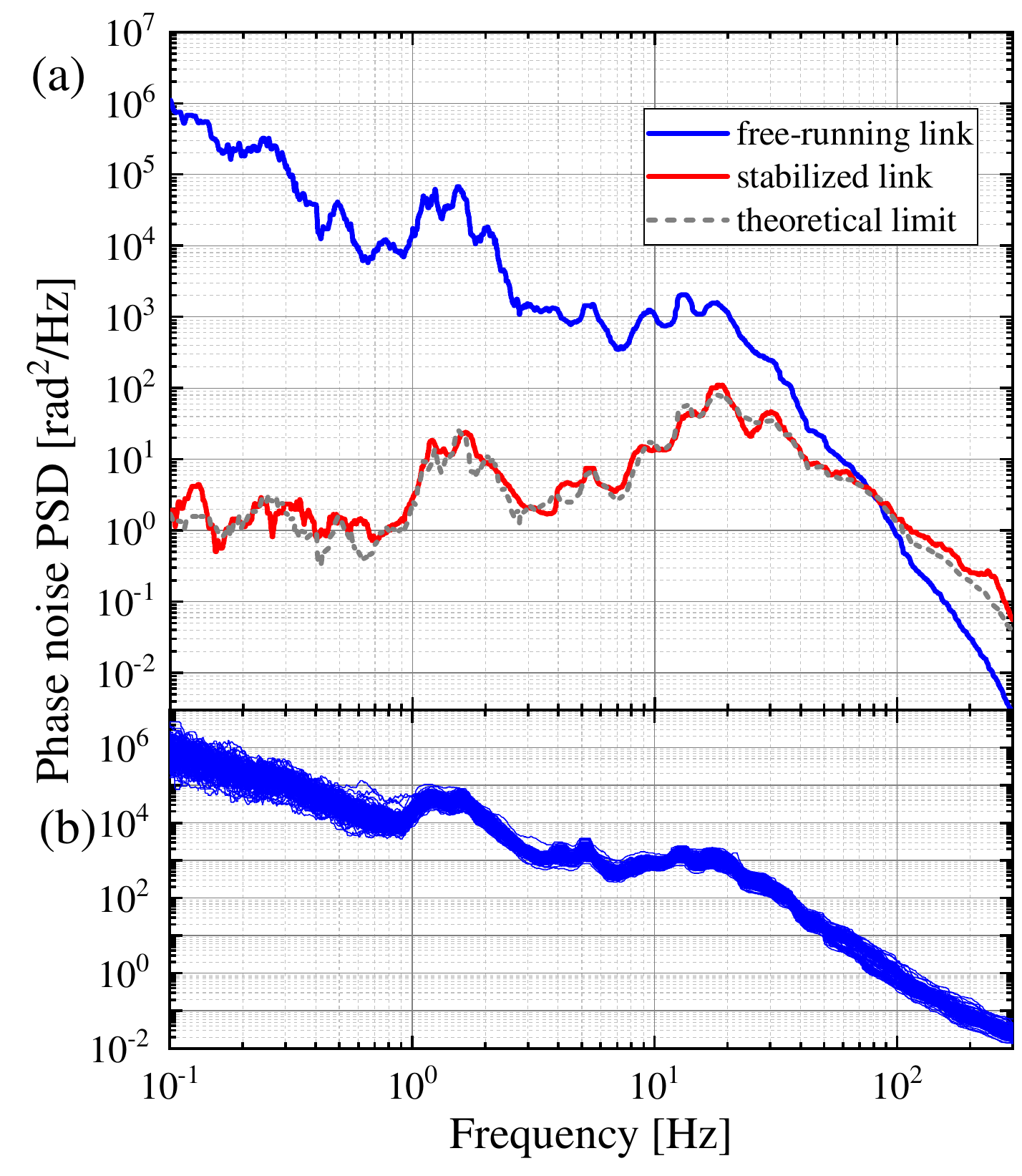}
\caption{(a) Measured phase noise power densities (PSDs) for the running 260 km urban fiber link (blue curve), the stabilized fiber link (red curve ) and the theoretical prediction (gray curve) (b) Measured phase noise PSDs of the 260 km free-running fiber for a whole day. The average phase jitter integrated from 0.1Hz to 100 kHz of this free-running 260km fiber link is about 240 rad.}
\label{fig4}
\end{figure}
Figure \ref{fig4}(a) shows the measured PSDs of the  free-running (blue curve) and stabilized (red curve) 260 km urban fiber link. The frequency resolution bandwidth is set as 1 mHz, resulting in 11 minutes for a single measurement.  The free-running phase noise PSD falls off approximately as $10000/f^2$ between 0.1 Hz and 10 Hz, the $f$ stands for the Fourier frequency. The results shows that the long-haul fiber link performs a higher level of phase noise per-unit-length at the 1 Hz (about 38.5 rad$^2$/Hz/km), which is much larger than the fiber noise observed in previous reports, such as $4$ rad$^2$/Hz/km (USA) \cite{williams2008high}, $0.5$ rad$^2$/Hz/km (Germany) \cite{raupach2014optical, raupach2015brillouin}, $3$ rad$^2$/Hz/km (France) \cite{Chiodo:15, lopez2010cascaded}, $20$ rad$^2$/Hz/km (Italy) \cite{calonico2014high}, $20$ rad$^2$/Hz/km (Japan) \cite{akatsuka2020optical}.  Furthermore, the phase noise PSD of this free-running fiber link exists obvious broad bumps around 1.4 Hz and 15 Hz. The peak variations around these two frequencies appear to be related to human activities, we can reasonably attribute this feature to infra-sound perturbations arising from urban traffic \cite{Raupach:14}.  To further characterize the free-running link with the different times, we measured phase noise PSDs of the 260 km free-running fiber for a whole day. The measured total 138 curves are shown in Fig. \ref{fig4} (b). We can observe, the phase noise at the frequency range ($f<1$ Hz) will fluctuate more than one order of magnitude. We attribute the high phase noise to the fact that this 260 km fiber link is laying underground along a noisy urban area. The measured passively stabilized phase noise PSD (red curve) also shown in Fig. \ref{fig4}(a), which is in good agreement with the PNC theoretical limit (gray dash line) as indicated in Eq. \ref{eq7}. More importantly, the strong servo bumps which existing in the ANC technique is disappeared, resulting in the reduction of the integrate timing jitter of the transferred light \cite{hu2020passive1, hu2021all,hu2020multinode}.

\subsection{Cycle-slip free optical frequency transfer}
\begin{figure}[htbp]
\centering
\includegraphics[width=1\linewidth]{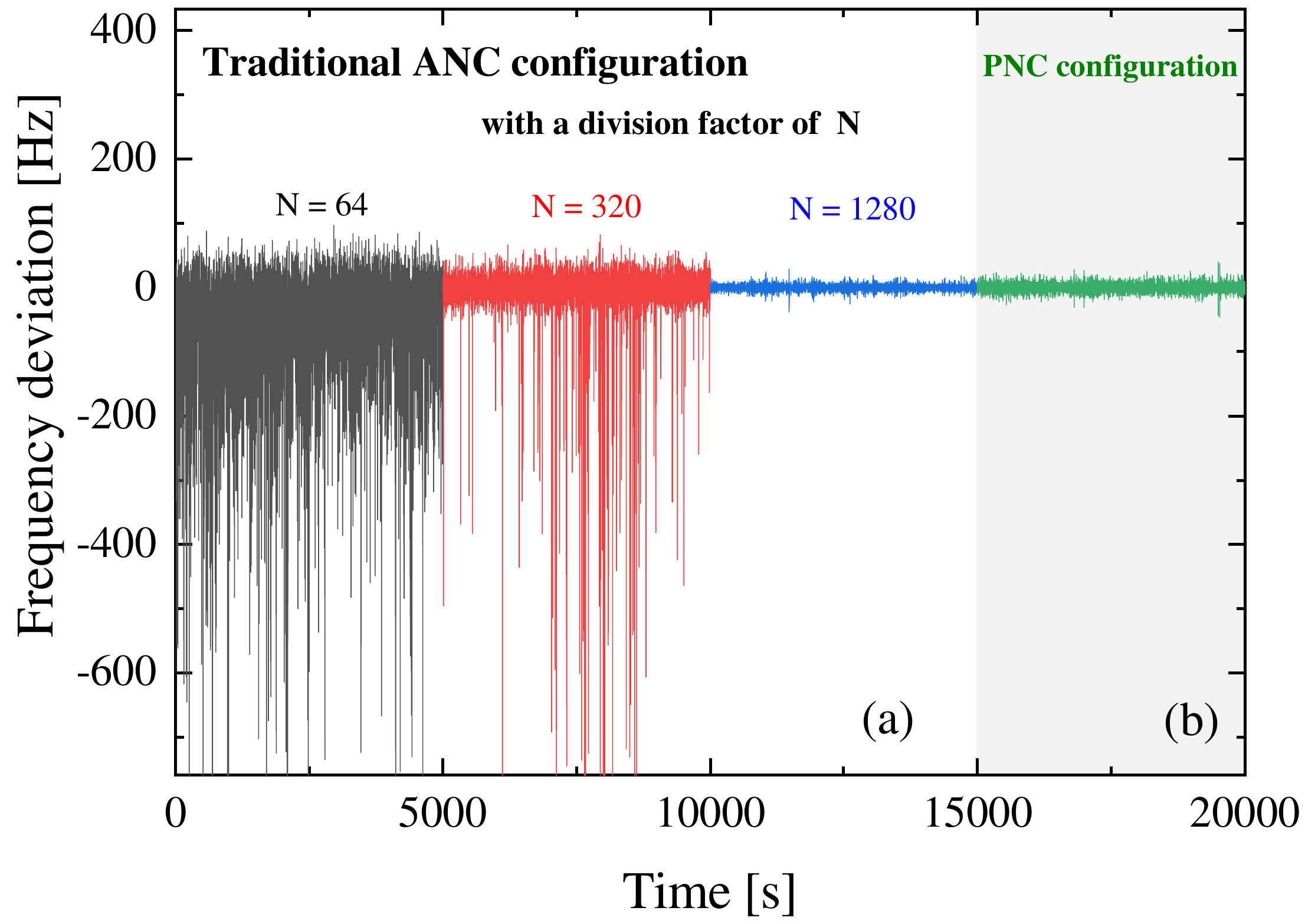}
\caption{Temporal behavior of the end-to-end frequency values of the stabilized 260 km urban fiber link respectively with the traditional ANC configuration (a) and the PNC configuration (b). }
\label{fig2}
\end{figure}

To examine the phase noise rejection capability of the  ANC and PNC techniques, we measured end-to-end frequencies in these two configurations. The corresponding results are shown in Fig.\ref{fig2}. In the ANC configuration, the angular frequency of $E_3$ indicated in Eq.\ref{eq4} is divided by a programmable frequency divider and sent to a phase comparator. The rule of thumb for the frequency dividing ratio is that the integrated phase jitter of the RF signal after frequency divider has to be much less than 1 rad. As shown in Fig.\ref{fig2}(a), the optical frequency transfer system couldn't perform properly when the frequency division ratio is 64. By increasing the ratio to 320, the system could be successfully locked, but there still exist many occasional cycle-slips because the free-running phase noise of this fiber link is not stationary as shown in Fig.\ref{fig4}(b). The  system could operate properly when the division factor is increased to 1280. Fig.\ref{fig2}(b) shows the temporal behavior of the end-to-end frequencies in the passively stabilized fiber link, demonstrating that the PNC configuration could run properly without the clear indication of the cycle-slips. As can be seen from the above experimental results, the active system needs a reasonable design for the frequency dividing ratio and the loop parameters, while the PNC-based optical frequency transfer system is  more efficient, simple, and robust employed in a noisy fiber link.

\subsection{Demonstration of the time base free optical frequency transfer}
\begin{figure}[htp]
\centering
\includegraphics[width=0.9\linewidth]{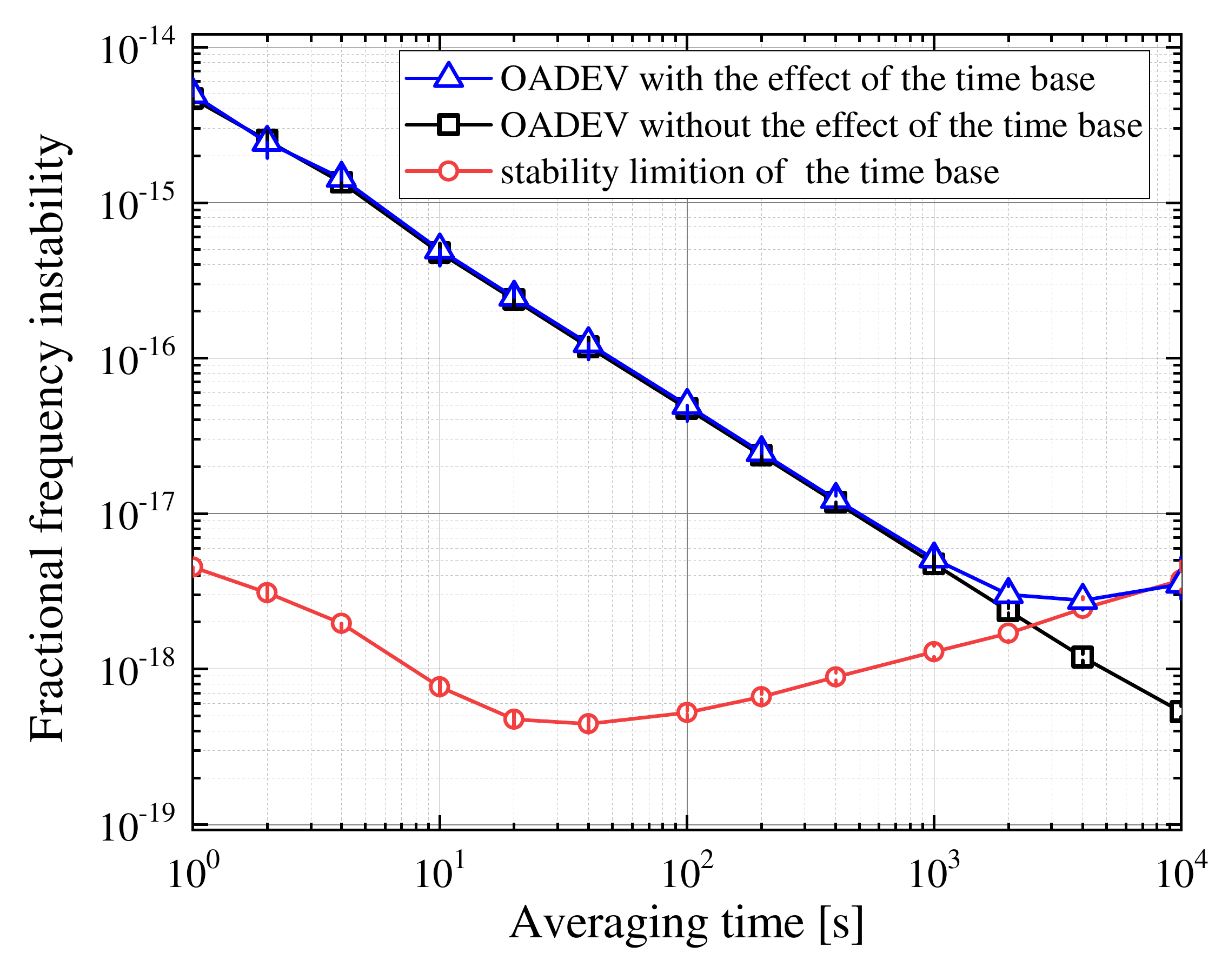}
\caption{Measured OADEV with (blue triangles) and without (black squares) the effect of the time base. The time base is from a quartz oscillator. To clearly illustrate the effect of the oscillator, the stability limited by the oscillator is also shown in red circles.}
\label{fig3}
\end{figure}
To illustrate the effect of the time base at the local site\cite{ma1994delivering, droste2013optical, calonico2014high, hu2020passive1, hu2021all}, before {feeding} the RF signal  onto the AOM1 we mix it {with} a RF signal $E_s$with the angular frequency of $\omega_s$. This measurement is based on the 150 km spooled fiber link.  The time base from a quartz oscillator converting to the optical frequency shows a instability of $4.5\times10^{-18}$ at 1 s and $2.9\times10^{-18}$ at 10,000 s in terms of the OADEV as illustrated the red circle markers in Fig. \ref{fig3}. We can clearly see that, once $E_s$ is employed in the system, the frequency of the fiber output light depends on the time base even when the PNC is properly working, as shown with the blue triangles in Fig.\ref{fig3}. The frequency instability is $4.9\times10^{-15}$ at 1 s averaging time, and is limited to the $10^{-18}$ level at the long averaging time due to the effect of the time base. Once $E_s$ is removed, the relative frequency instability is $4.9\times10^{-15}$ at 1 s averaging time and scales down to $5.1\times10^{-19}$ at 10,000 s averaging time (black squares), independent of the time base at the local site. We can conclude that with the implementation of the proposed technique, the effect of the time base can be effectively avoided.


\subsection{Optical frequency transfer stability characterization}

\begin{figure}[h]
\centering
\includegraphics[width=0.95\linewidth]{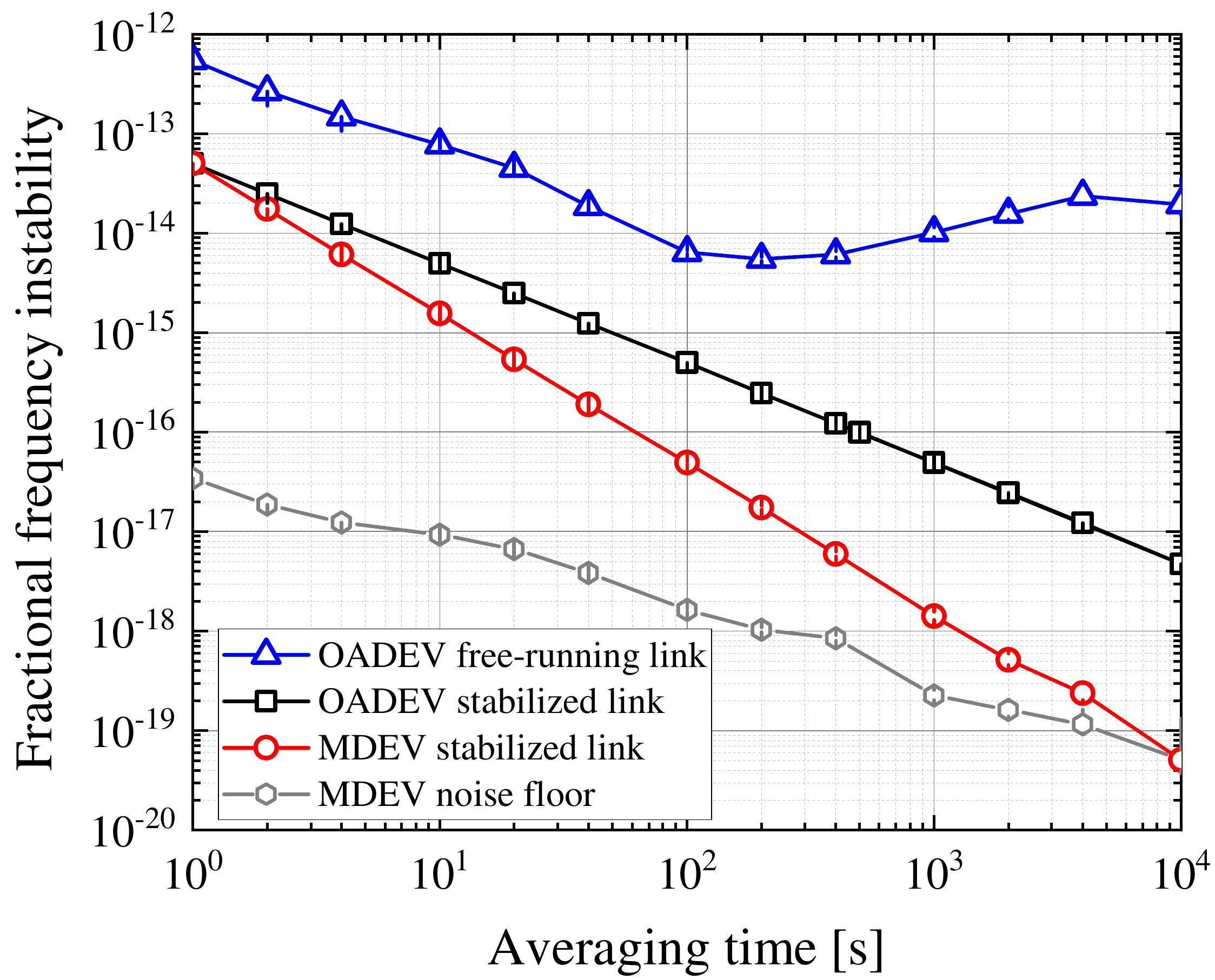}
\caption{Measured fractional frequency instability of the stabilized 260 km urban fiber link in terms of the OADEV (black squares) and the MDEV (red circles) with a $\Pi$-type frequency counter. Measured fractional frequency instability of the free-running fiber link (blue triangles) and noise floor of this fiber-based interferometer are also shown (gray hexagons).}
\label{fig5}
\end{figure}

Complementary to the frequency-domain feature, the time-domain characterization of the corresponding fractional frequency instability in the 260 km noisy urban fiber link is shown in Fig. \ref{fig5}. The fractional frequency instability of the end-to-end frequency measurements are expressed in terms of the the OADEV (black squares) and modified Allan deviation (MDEV, red circles). With the implementation of the PNC technique on the 260 km urban fiber link, we demonstrate a stability of $4.9\times10^{-14}$ at 1 s integration time and scales down to $4.7\times10^{-18}$ at the integration time of 10,000 s for the OADEV with a slope of $\tau^{-1}$, while the MDEV for the same frequency data decreases to $5.1\times10^{-20}$ at 10,000 s as a slope of $\tau^{-3/2}$. The results indicate that the white phase noise is dominated in the phase-compensated fiber link. In our experiment, we observe that the stability of optical frequency transfer is improved by three orders of magnitude at the integration time of 10,000 s by activicating the PNC setup.

\begin{figure}[htbp]
\centering
\includegraphics[width=0.95\linewidth]{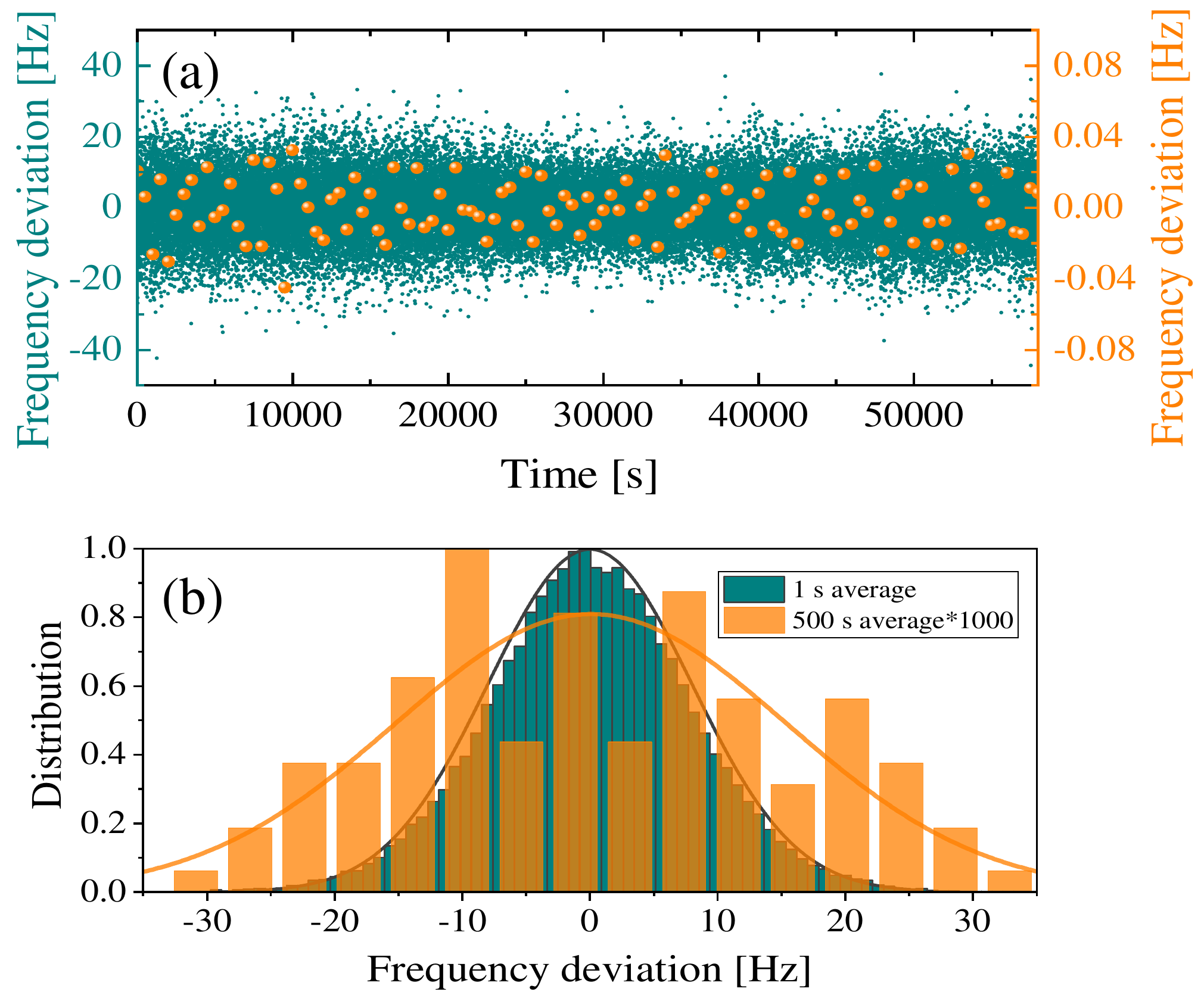}
\caption{(a) Frequency comparison between sent and transferred optical frequency through the 260 km urban fiber. The 58,000 data points were taken with a dead-time free $\Pi$-type frequency counters with a gate time of 1 s (dark cyan points, left axis). The unweighted arithmetic means for all cycle-slip free 500 intervals have been calculated, resulting in 116 data points (orange dots, right axis). {(b) Histograms and Gaussian fits for the obtained frequency values. A relative difference between sent and transferred frequency of $(0.41\pm4.7)\times10^{-18}$ is calculated.}}
\label{fig6}
\end{figure}

\subsection{Frequency transfer accuracy analysis}

To reveal the existing frequency offset of the transmitted light which may not be disclosed in the instability analysis, we further perform the evaluation of the optical frequency transfer accuracy. Figure \ref{fig6}(a) shows the frequency deviation of the beat-note’s data, recorded with a 1 s gate time and the total 58000 data (left axis). By averaging the data every 500 seconds, resulting  in 116 data points as shown in the right axis of Fig. \ref{fig6}. The total 116 data points have an arithmetic mean of 78.9 $\mu$Hz ($4.1\times10^{-19}$) and a standard deviation of 9.2 mHz ($4.7\times10^{-17}$), which is a factor of 500  smaller than the ADEV at 1s as expected for this $\Pi$-type evaluation. 
Considering the long-term stability of frequency transfer as illustrated in Fig. \ref{fig5}, we conservatively estimate the accuracy of the transmitted optical signal as shown in the last data point of the ADEV, resulting in a relative frequency accuracy of $4.7\times10^{-18}$. {We can conclude that there is no systematic frequency offset between the sent and transferred frequencies at a level of a few $10^{-18}$.}

\section{Discussion and Conclusion}
\label{sec6}

In summary, we for the first time demonstrated coherence transfer of an optical frequency through a 260 km-long urban fiber with the PNC technique without the effect of the time-base at the local site.  Neither precise phase detector with the large dynamic range nor phase locked loop are longer needed in this PNC-based optical frequency transfer system, lowering the probability of  cycle-slips for a noisy fiber link.  Moreover, getting rid of the time base effect maintains the coherence of the optical frequency reference as opposed in conventional active techniques, significantly enabling the simplification of the experimental setup.  {After being compensated,} the relative frequency transfer instability is $4.9\times10^{-14}$ at 1 s averaging time and  $5.1\times10^{-20}$ at 10,000 s averaging time in terms of the MDEV. The frequency uncertainty of the light after transferring through the fiber achieves $(0.41\pm4.7)\times10^{-18}$. Although the residual phase noise with the PNC technique is 7 times higher than that in the conventional ANC system, the phase noise could be suppressed by dividing the whole link into several sub-sections with laser repeater stations \cite{Chiodo:15, Lopez:12, Guillou-Camargo:18}. Although here we only demonstrate that the time base effect can be effectively avoided in the PNC technique based optical frequency transfer, it can also be used for the ANC-based scheme by slightly modifying the configuration of the setup. The proposed technique demonstrated here opens a way to a broad distribution of an ultrastable frequency reference with high coherence without any effects coming from the RF reference and enables a wide range of applications beyond metrology over fiber networks.


\ifCLASSOPTIONcaptionsoff
  \newpage
\fi



%
\bibliographystyle{IEEEtran}
\bibliography{Optics}

\begin{thebibliography}{10}
\providecommand{\url}[1]{#1}
\csname url@samestyle\endcsname
\providecommand{\newblock}{\relax}
\providecommand{\bibinfo}[2]{#2}
\providecommand{\BIBentrySTDinterwordspacing}{\spaceskip=0pt\relax}
\providecommand{\BIBentryALTinterwordstretchfactor}{4}
\providecommand{\BIBentryALTinterwordspacing}{\spaceskip=\fontdimen2\font plus
\BIBentryALTinterwordstretchfactor\fontdimen3\font minus
  \fontdimen4\font\relax}
\providecommand{\BIBforeignlanguage}[2]{{%
\expandafter\ifx\csname l@#1\endcsname\relax
\typeout{** WARNING: IEEEtran.bst: No hyphenation pattern has been}%
\typeout{** loaded for the language `#1'. Using the pattern for}%
\typeout{** the default language instead.}%
\else
\language=\csname l@#1\endcsname
\fi
#2}}
\providecommand{\BIBdecl}{\relax}
\BIBdecl

\bibitem{PhysRevLett.120.103201}
G.~E. Marti, R.~B. Hutson, A.~Goban, S.~L. Campbell, N.~Poli, and J.~Ye,
  ``Imaging optical frequencies with $100\text{ }\text{
  }\ensuremath{\mu}\mathrm{Hz}$ precision and $1.1\text{ }\text{
  }\ensuremath{\mu}\mathrm{m}$ resolution,'' \emph{Phys. Rev. Lett.}, vol. 120,
  p. 103201, Mar 2018.

\bibitem{schioppo2017ultrastable}
M.~Schioppo, R.~C. Brown, W.~F. McGrew, N.~Hinkley, R.~J. Fasano, K.~Beloy,
  T.~Yoon, G.~Milani, D.~Nicolodi, J.~A. Sherman \emph{et~al.}, ``Ultrastable
  optical clock with two cold-atom ensembles,'' \emph{Nat. Photonics}, vol.~11,
  no.~1, pp. 48--52, 2017.

\bibitem{McGrew:2018aa}
W.~F. McGrew, X.~Zhang, R.~J. Fasano, S.~A. Sch{\"a}ffer, K.~Beloy,
  D.~Nicolodi, R.~C. Brown, N.~Hinkley, G.~Milani, M.~Schioppo, T.~H. Yoon, and
  A.~D. Ludlow, ``Atomic clock performance enabling geodesy below the
  centimetre level,'' \emph{Nature}, vol. 564, no. 7734, pp. 87--90, 2018.

\bibitem{chou2010}
T.~R. C.~W.~Chou, D. B.~Hume and D.~J. Wineland, ``Optical clocks and
  relativity,'' \emph{Science}, vol. 329, pp. 1630---1633, 2010.

\bibitem{delva2017}
P.~Delva, J.~Lodewyck, S.~Bilicki, E.~Bookjans, G.~Vallet, R.~Le~Targat, P.-E.
  Pottie, C.~Guerlin, F.~Meynadier, C.~Le~Poncin-Lafitte, O.~Lopez,
  A.~Amy-Klein, W.-K. Lee, N.~Quintin, C.~Lisdat, A.~Al-Masoudi, S.~D\"orscher,
  C.~Grebing, G.~Grosche, A.~Kuhl, S.~Raupach, U.~Sterr, I.~R. Hill, R.~Hobson,
  W.~Bowden, J.~Kronj\"ager, G.~Marra, A.~Rolland, F.~N. Baynes, H.~S.
  Margolis, and P.~Gill, ``Test of special relativity using a fiber network of
  optical clocks,'' \emph{Phys. Rev. Lett.}, vol. 118, p. 221102, Jun 2017.

\bibitem{parker2018measurement}
R.~H. Parker, C.~Yu, W.~Zhong, B.~Estey, and H.~M{\"u}ller, ``Measurement of
  the fine-structure constant as a test of the standard model,''
  \emph{Science}, vol. 360, no. 6385, pp. 191--195, 2018.

\bibitem{Roberts_2020}
B.~M. Roberts, P.~Delva, A.~Al-Masoudi, A.~Amy-Klein, C.~B{\ae}rentsen,
  C.~F.~A. Baynham, E.~Benkler, S.~Bilicki, S.~Bize, W.~Bowden, J.~Calvert,
  V.~Cambier, E.~Cantin, E.~A. Curtis, S.~D{\"o}rscher, M.~Favier, F.~Frank,
  P.~Gill, R.~M. Godun, G.~Grosche, C.~Guo, A.~Hees, I.~R. Hill, R.~Hobson,
  N.~Huntemann, J.~Kronj{\"a}ger, S.~Koke, A.~Kuhl, R.~Lange, T.~Legero,
  B.~Lipphardt, C.~Lisdat, J.~Lodewyck, O.~Lopez, H.~S. Margolis,
  H.~{\'{A}}lvarez-Mart{\'{\i}}nez, F.~Meynadier, F.~Ozimek, E.~Peik, P.-E.
  Pottie, N.~Quintin, C.~Sanner, L.~D. Sarlo, M.~Schioppo, R.~Schwarz,
  A.~Silva, U.~Sterr, C.~Tamm, R.~L. Targat, P.~Tuckey, G.~Vallet,
  T.~Waterholter, D.~Xu, and P.~Wolf, ``Search for transient variations of the
  fine structure constant and dark matter using fiber-linked optical atomic
  clocks,'' \emph{New Journal of Physics}, vol.~22, no.~9, p. 093010, sep 2020.

\bibitem{grotti2018geodesy}
J.~Grotti, S.~Koller, S.~Vogt, S.~H{\"a}fner, U.~Sterr, C.~Lisdat, H.~Denker,
  C.~Voigt \emph{et~al.}, ``Geodesy and metrology with a transportable optical
  clock,'' \emph{Nat. Phys.}, vol.~14, no.~5, pp. 437--441, 2018.

\bibitem{kolkowitz2016gravitational}
S.~Kolkowitz, I.~Pikovski, N.~Langellier, M.~D. Lukin, R.~L. Walsworth, and
  J.~Ye, ``Gravitational wave detection with optical lattice atomic clocks,''
  \emph{Phys. Rev. D}, vol.~94, no.~12, p. 124043, 2016.

\bibitem{graham2013new}
P.~W. Graham, J.~M. Hogan, M.~A. Kasevich, and S.~Rajendran, ``New method for
  gravitational wave detection with atomic sensors,'' \emph{Phys. Rev. Lett.},
  vol. 110, no.~17, p. 171102, 2013.

\bibitem{hu2017atom}
L.~Hu, N.~Poli, L.~Salvi, and G.~M. Tino, ``Atom interferometry with the {S}r
  optical clock transition,'' \emph{Phys. Rev. Lett.}, vol. 119, no.~26, p.
  263601, 2017.

\bibitem{takamoto2005optical}
M.~Takamoto, F.-L. Hong, R.~Higashi, and H.~Katori, ``An optical lattice
  clock,'' \emph{Nature}, vol. 435, no. 7040, pp. 321--324, 2005.

\bibitem{ma1994delivering}
L.-S. Ma, P.~Jungner, J.~Ye, and J.~L. Hall, ``Delivering the same optical
  frequency at two places: accurate cancellation of phase noise introduced by
  an optical fiber or other time-varying path,'' \emph{Opt. Lett.}, vol.~19,
  no.~21, pp. 1777--1779, 1994.

\bibitem{droste2013optical}
S.~Droste, F.~Ozimek, T.~Udem, K.~Predehl, T.~H{\"a}nsch, H.~Schnatz,
  G.~Grosche, and R.~Holzwarth, ``Optical-frequency transfer over a single-span
  1840 km fiber link,'' \emph{Physical review letters}, vol. 111, no.~11, p.
  110801, 2013.

\bibitem{calonico2014high}
D.~Calonico, E.~Bertacco, C.~Calosso, C.~Clivati, G.~Costanzo, M.~Frittelli,
  A.~Godone, A.~Mura, N.~Poli, D.~Sutyrin \emph{et~al.}, ``High-accuracy
  coherent optical frequency transfer over a doubled 642-km fiber link,''
  \emph{Appl. Phys. B}, vol. 117, no.~3, pp. 979--986, 2014.

\bibitem{Chiodo:15}
N.~Chiodo, N.~Quintin, F.~Stefani, F.~Wiotte, E.~Camisard, C.~Chardonnet,
  G.~Santarelli, A.~Amy-Klein, P.-E. Pottie, and O.~Lopez, ``Cascaded optical
  fiber link using the internet network for remote clocks comparison,''
  \emph{Opt. Express}, vol.~23, no.~26, pp. 33\,927--33\,937, Dec 2015.

\bibitem{redpitaya2018}
A.~Tourigny-Plante, V.~Michaud-Belleau, N.~Bourbeau~H{\'e}bert, H.~Bergeron,
  J.~Genest, and J.-D. Desch{\^e}nes, ``An open and flexible digital
  phase-locked loop for optical metrology,'' \emph{Review of Scientific
  Instruments}, vol.~89, no.~9, p. 093103, 2018.

\bibitem{williams2008high}
P.~A. Williams, W.~C. Swann, and N.~R. Newbury, ``High-stability transfer of an
  optical frequency over long fiber-optic links,'' \emph{J. Opt. Soc. Am. B},
  vol.~25, no.~8, pp. 1284--1293, 2008.

\bibitem{raupach2014optical}
S.~Raupach, A.~Koczwara, and G.~Grosche, ``Optical frequency transfer via a 660
  km underground fiber link using a remote brillouin amplifier,'' \emph{Optics
  express}, vol.~22, no.~22, pp. 26\,537--26\,547, 2014.

\bibitem{lopez2010cascaded}
O.~Lopez, A.~Haboucha, F.~K{\'e}f{\'e}lian, H.~Jiang, B.~Chanteau, V.~Roncin,
  C.~Chardonnet, A.~Amy-Klein, and G.~Santarelli, ``Cascaded multiplexed
  optical link on a telecommunication network for frequency dissemination,''
  \emph{Opt. Express}, vol.~18, no.~16, pp. 16\,849--16\,857, 2010.

\bibitem{akatsuka2020optical}
T.~Akatsuka, T.~Goh, H.~Imai, K.~Oguri, A.~Ishizawa, I.~Ushijima, N.~Ohmae,
  M.~Takamoto, H.~Katori, T.~Hashimoto \emph{et~al.}, ``Optical frequency
  distribution using laser repeater stations with planar lightwave circuits,''
  \emph{Optics express}, vol.~28, no.~7, pp. 9186--9197, 2020.

\bibitem{hu2020passive1}
L.~Hu, X.~Tian, G.~Wu, and J.~Chen, ``Passive optical phase noise
  cancellation,'' \emph{Opt. Lett.}, vol.~45, no.~15, pp. 4308--4311, Aug 2020.

\bibitem{hu2021all}
L.~Hu, R.~Xue, X.~Tian, G.~Wu, and J.~Chen, ``All-passive multiple-place
  optical phase noise cancellation,'' \emph{Optics Letters}, vol.~46, no.~6,
  pp. 1381--1384, 2021.

\bibitem{Lopez:12}
O.~Lopez, A.~Haboucha, B.~Chanteau, C.~Chardonnet, A.~Amy-Klein, and
  G.~Santarelli, ``Ultra-stable long distance optical frequency distribution
  using the internet fiber network,'' \emph{Opt. Express}, vol.~20, no.~21, pp.
  23\,518--23\,526, 2012.

\bibitem{deng2016}
X.~{Deng}, J.~{Liu}, D.-D. {Jiao}, J.~{Gao}, Q.~{Zang}, G.-J. {Xu}, R.-F.
  {Dong}, T.~{Liu}, and S.-G. {Zhang}, ``Coherent transfer of optical frequency
  over 112 km with instability at the 10$^{-20}$ level,'' \emph{Chinese Physics
  Letters}, vol.~33, no.~11, pp. 114\,202--1\,142\,056, Nov 2016.

\bibitem{cheng2005}
C.-L. Cheng, F.-R. Chang, and K.-Y. Tu, ``Highly accurate real-time gps carrier
  phase-disciplined oscillator,'' \emph{IEEE Transactions on Instrumentation
  and Measurement}, vol.~54, no.~2, pp. 819--824, 2005.

\bibitem{Wu16}
L.~Wu, Y.~Jiang, C.~Ma, H.~Yu, Z.~Bi, and L.~Ma, ``Coherence transfer of
  subhertz-linewidth laser light via an optical fiber noise compensated by
  remote users,'' \emph{Opt. Lett.}, vol.~41, no.~18, pp. 4368--4371, Sep 2016.

\bibitem{stefani2015tackling}
F.~Stefani, O.~Lopez, A.~Bercy, W.-K. Lee, C.~Chardonnet, G.~Santarelli, P.-E.
  Pottie, and A.~Amy-Klein, ``Tackling the limits of optical fiber links,''
  \emph{JOSA B}, vol.~32, no.~5, pp. 787--797, 2015.

\bibitem{hu2020multinode}
L.~{Hu}, X.~{Tian}, G.~{Wu}, M.~{Kong}, J.~{Shen}, and J.~{Chen}, ``Multi-node
  optical frequency dissemination with post automatic phase correction,''
  \emph{J. Light. Technol.}, vol.~38, no.~14, pp. 3644--3651, 2020.

\bibitem{hu2021branching}
R.~Xue, L.~Hu, J.~Shen, J.~Chen, and G.~Wu, ``Branching optical frequency
  transfer with enhanced post automatic phase noise cancellation,''
  \emph{Journal of Lightwave Technology}, pp. 1--1, 2021.

\bibitem{raupach2015brillouin}
S.~M. Raupach, A.~Koczwara, and G.~Grosche, ``Brillouin amplification supports
  1$\times$ 10- 20 uncertainty in optical frequency transfer over 1400 km of
  underground fiber,'' \emph{Physical Review A}, vol.~92, no.~2, p. 021801,
  2015.

\bibitem{Raupach:14}
S.~M.~F. Raupach, A.~Koczwara, and G.~Grosche, ``Optical frequency transfer via
  a 660 km underground fiber link using a remote brillouin amplifier,''
  \emph{Opt. Express}, vol.~22, no.~22, pp. 26\,537--26\,547, Nov 2014.

\bibitem{Guillou-Camargo:18}
F.~Guillou-Camargo, V.~M\'{e}noret, E.~Cantin, O.~Lopez, N.~Quintin,
  E.~Camisard, V.~Salmon, J.-M.~L. Merdy, G.~Santarelli, A.~Amy-Klein, P.-E.
  Pottie, B.~Desruelle, and C.~Chardonnet, ``First industrial-grade coherent
  fiber link for optical frequency standard dissemination,'' \emph{Appl. Opt.},
  vol.~57, no.~25, pp. 7203--7210, Sep 2018.

\end{thebibliography}


\end{document}